\documentclass[12pt]{extarticle}
\usepackage{setspace}

\usepackage{setspace}
\usepackage[T1]{fontenc}
\usepackage{times}
\usepackage{palatino} 
\usepackage{lmodern}
\usepackage[latin1]{inputenc}
\usepackage{epsfig}
\usepackage[english]{babel}
\usepackage{color}
\usepackage{graphicx}
\usepackage{dcolumn}
\usepackage{moreverb}
\usepackage{amsmath,amssymb,amsfonts}
\usepackage[all]{xy}

\usepackage[svgnames]{xcolor}
\usepackage{tikz}

\def\nudge{.5}

\tikzset{axis/.style={ultra thick, Red!75!black, -latex, shorten <=-\nudge cm, shorten >=-2*\nudge cm}}
\tikzset{line/.style={thick,Green}}

\begin{document}
\numberwithin{equation}{section}
\newcommand{\boxedeqn}[1]{%
  \[\fbox{%
      \addtolength{\linewidth}{-2\fboxsep}%
      \addtolength{\linewidth}{-2\fboxrule}%
      \begin{minipage}{\linewidth}%
      \begin{equation}#1\end{equation}%
      \end{minipage}%
    }\]%
}


\newsavebox{\fmbox}
\newenvironment{fmpage}[1]
     {\begin{lrbox}{\fmbox}\begin{minipage}{#1}}
     {\end{minipage}\end{lrbox}\fbox{\usebox{\fmbox}}}

\raggedbottom
\onecolumn

\parindent 8pt
\parskip 10pt
\baselineskip 16pt
\noindent\title*{{\LARGE{\textbf{On superintegrable monopole systems }}}}
\newline
\newline
\newline
Md Fazlul Hoque, Ian Marquette and Yao-Zhong Zhang
\newline
\newline
 School of Mathematics and Physics, The University of Queensland, Brisbane, QLD 4072, Australia
\newline
\newline
E-mail: m.hoque@uq.edu.au; i.marquette@uq.edu.au; yzz@maths.uq.edu.au
\newline
\newline
\begin{abstract}
Superintegrable systems  with monopole interactions in flat and curved spaces have attracted much attention. For example, models in spaces with a Taub-NUT metric are well-known to admit the Kepler-type symmetries and provide non-trivial generalizations of the usual Kepler problems. In this paper, we overview new families of superintegrable Kepler, MIC-harmonic oscillator and deformed Kepler systems interacting with Yang-Coulomb monopoles in the flat and curved Taub-NUT spaces. We present their higher-order, algebraically independent integrals of motion via the direct and constructive approaches which prove the superintegrability of the models. The integrals form symmetry polynomial algebras of the systems with structure constants involving Casimir operators of certain Lie algebras. Such algebraic approaches provide a deeper understanding to the degeneracies of the energy spectra and connection between wave functions and differential equations and geometry. 
\end{abstract}

\section{Introduction}

Classical and quantum superintegrable Hamiltonian systems with non-scalar potentials are significant research topics in mathematical physics. Well-known examples are the Kepler problems with additional magnetic monopole interactions known as MICZ-Kepler monopole systems \cite{Mci1, Zwa1}.

The family of Taub-NUT metrics provide interesting models for investigating classical and quantum conserved quantities on curved spaces. The geodesic of the generalized Taub-NUT metrics properly explains the motion of well-separated monopole-monopole interactions which provide non-trivial generalizations of the Kepler and harmonic oscillator systems \cite{Grs1, Gib1, Feh1, Man1, Lab1, Gro1, Iwa2, Iwa1, Mar2}. The 5D Kepler systems interacting with $su(2)$ monopoles or Yang-Coulomb monopoles have been investigated in \cite{Sis1, Mad1, Mad2, Ple1, Mad5, Ple2, Kal1, Ner1, Mad12, Kar1, Bel2}.  Recently, there is a lot of international research interest in generalizing such monopole models to higher dimensions via different approaches \cite{Mar2, Mad3, Mad4, Men1, Ran1, Mar1}.  

Quadratic algebras are one important class of symmetry algebras for certain superintegrable systems \cite{Gra1}. The quadratic algebra $Q(3)$ with three generators given by 2nd-order integrals of motion of 3D superintegrable models and its realizations in term of deformed oscillators have been presented in \cite{Das1} by Daskaloyannis. We extended this work to $N$-dimensional superintegrable quantum systems with non-central scalar potentials \cite{FH1, FH2}. It was shown that the symmetry algebras of the $N$-dimensional superintegrable models have the intrinsic form $Q(3)\oplus L_1\oplus L_2\oplus\dots$, where $L_1, L_2,\dots$ are certain Lie algebras \cite{FH7}. We also applied the recurrence approach to the $N$-dimensional superintegrable systems and constructed the corresponding higher-rank polynomial algebras \cite{FH3}. 
Recently, we applied the direct and recurrence approaches to superintegrable models with monopole interactions in both the flat and curved Taub-NUT spaces \cite{FH4, FH5, FH6}.

In this paper, we review some of the main results obtained in the three papers \cite{FH4, FH5, FH6} and show the monopole systems have the similar algebra structure $P(3)\oplus\mathcal{L}$ as their symmetry algebra, where $P(3)$ is a polynomial algebra involving three generators and $\mathcal{L}=\mathcal{L}_1\oplus\mathcal{L}_2,\dots$ is a direct sum of certain Lie algebras $\mathcal{L}_1, \mathcal{L}_2\dots$. Such algebraic structure allowed us to give purely algebraic derivations for the complete energy spectra of the models.

\section{Superintegrable monopole models}
In this section, we present the new families of Hamiltonian models interacting with monopoles in the flat and curved Taub-NUT spaces which were introduced in our recent papers \cite{FH4, FH5, FH6}.

Consider the generalized Taub-NUT metric in 3D \cite{Grs1}
\begin{eqnarray}
ds^2=f(r)d\textbf{r}^2+g(r)(d\psi+\textbf{A}. d\textbf{r} )^2,\label{1mc1}
\end{eqnarray}
where $r=\sqrt{x_0^2+x_1^2+x_2^2}$, $f(r)$ and $g(r)$ are certain functions of $r$,  $\psi$ is the additional angular variable, $d\textbf{r}^2=dx_0^2+dx_1^2+dx_2^3$ is the 3D line element and $\textbf{A}$ is the vector potential for the magnetic field.
The following two monopole systems in the 3D Taub-NUT space were introduced in \cite{FH4, FH6}
\begin{eqnarray}
&&H_{K}=\frac{1}{2}\left[\frac{1}{f(r)}\left\{\textbf{p}^2+\frac{c_0}{2r}+c_4\right\}+\frac{Q^2}{g(r)}\right]+\frac{1}{4f(r)}\left\{\frac{c_2}{r(r+x_0)}+\frac{c_3}{r(r-x_0)}\right\},\nonumber\\&&\label{k1}
\\&&
H_{MIC}=\frac{1}{2}\left[\frac{1}{f(r)}\left\{\textbf{p}^2+\frac{c_0 r^2}{2}+c_4\right\}+\frac{Q^2}{g(r)}\right]\label{kf1},
\end{eqnarray}
where $p_i=-i(\partial_i-i A_i Q)$,  $Q=-i\partial_\psi$ and $A_1=0$, $A_2=\frac{x_3}{r(r+x_0)}$, $A_3=\frac{-x_1}{r(r+x_0)}$ are the components of $\textbf{A}$ which are identified with the monopole interactions; $f(r)=\frac{a_1}{r}+b_1$, $g(r)=\frac{r(a_1+b_1 r)}{1+c_1 r+d_1 r^2}$ for model $H_k$, $f(r)=a_1 r^2+b_1$, $g(r)=\frac{r^2(a_1r^2+b_1)}{1+c_1 r^2+d_1 r^4}$ for model $H_{MIC}$, and $a_1$, $b_1$, $c_0$, $c_1$, $c_2$, $c_3$, $c_4$, $d_1$ are constants.
 The operators $p_i$,  $Q$ obey the commutation relations
\begin{eqnarray}
[p_i,p_j]=i\epsilon_{ijk}B_k Q, \quad [p_i,Q]=0,\quad \textbf{B}=\frac{\textbf{r}}{r^3}.\label{kpf1}
\end{eqnarray} 
The systems (\ref{k1}) and (\ref{kf1}) are generalized Hartmann systems \cite{Hartm1} in the curved Taub-NUT space with abelian monopole interactions. System (\ref{k1}) is referred to as Kepler monopole system which contains the Kaluza-Klein \cite{Gib1, Mar1} and MICZ monopole system \cite{Mci1, Zwa1} as special cases. System (\ref{kf1}) is referred to as MIC-harmonic oscillator monopole system.

We also consider the monopole model in the 5D flat Euclidean space \cite{FH5}, 
\begin{eqnarray}
&&H_{YC}=\frac{1}{2}\left(\pi_j \right)^{2}  +\frac{\hbar^{2}}{2 r^{2}} \hat{T}^2 -\frac{c_{0}}{r} + \frac{c_{1}}{r(r+x_{0})} + \frac{c_{2}}{r(r-x_{0})},\label{KP1}
\end{eqnarray}
where $\pi_j= -i\hbar \frac{\partial}{\partial x_{j}} -\hbar A_{j}^{a} T_{a}$, $j=0, 1, 2, 3, 4$, $c_1$, $c_2$ are positive real constants and $\{T_a, a=1, 2, 3\}$ are the $su(2)$ generators which satisfy $[T_a, T_b]=i\epsilon_{abc}T_c$. $\hat{T}^2=T^a T^a$ is the Casimir operator of $su(2)$ and $A^a_j$ are the components of the monopole potential $\textbf{A}^a$ of the form
\begin{eqnarray}
A_j^a=\frac{2i}{r(r+x_0)}\tau^a_{jk}x_k.
\end{eqnarray}
Here $\tau^a$ are the $5\times 5$ matrices
\begin{eqnarray*}
\tau^1 =\frac{1}{2}
\begin{pmatrix}
0 & 0 & 0 \\
0 & 0 & -i\sigma^1 \\
0 & i\sigma^1 & 0
\end{pmatrix},\quad
\tau^2 =\frac{1}{2}
\begin{pmatrix}
0 & 0 & 0 \\
0 & 0 & i\sigma^3 \\
0 & -i\sigma^3 & 0
\end{pmatrix},\quad
\tau^3 =\frac{1}{2}
\begin{pmatrix}
0 & 0 & 0 \\
0 & \sigma^2 & 0 \\
0 & 0 &  \sigma^2
\end{pmatrix},
\end{eqnarray*}
which satisfy $[\tau^a, \tau^b]=i\epsilon_{abc}\tau^c$ and $\sigma^i$ are the Pauli matrices.
The system (\ref{KP1}) is referred to as the 5D Kepler system in Yang-Coulomb monopole field with non-central potentials.

\section{Integrals of motion and algebra structures}
In this section, we highlight the common feature of the algebra structures generated by the integrals of motion of the superintegrable monopole models (\ref{k1}), (\ref{kf1}) and (\ref{KP1}).

Consider a superintegrable system $H$ which allows two additional second-order integrals of motion $A$ and $B$ and other first-order integrals of motion $\{F_i\}$. Suppose that $A$, $B$ and $F_i$ form the quadratic algebra structure $Q(3)\oplus \mathcal{L}$ \cite{Das1, FH4, FH5},
\begin{eqnarray}
\left.\begin{aligned}
 &[A,B]=C,
\\&
 [A,C]=\gamma\{A,B\}+\epsilon B+\zeta(H,G_i),
\\&
[B,C]=-\gamma B^2+d(H,G_i)A+z(H,G_i)\end{aligned}\right\}\label{QA1}
\end{eqnarray}
\begin{eqnarray}
[F_i,F_j]=\sum_k C_{ij}^k F_k, 
\end{eqnarray}
where $\gamma, \epsilon$ are constants, $\zeta, d, z$ are functions of $H, G_i$ with $\{G_i\}$ being the central elements of $Q(3)\oplus \mathcal{L}$, $\mathcal{L}=\mathcal{L}_1\oplus\dots\oplus\mathcal{L}_n$ is a semi-simple Lie algebra and $C_{ij}^k$ are the structure constants of $\mathcal{L}$. Thus the structure of the full symmetry algebra for the Hamiltonian $H$ is a direct sum $Q(3)\oplus \mathcal{L}$ of the quadratic algebra $Q(3)$ and the semi-simple Lie algebra $\mathcal{L}$.

Let $K$ be the Casimir operator of this quadratic algebra. Then $K$ commutes with all generators of the algebra, that is, $[K,A]=[K,B]=0=[K,C]$, and it has the form \cite{Das1, FH4, FH5}
\begin{eqnarray}
K=C^2-\gamma\{A,B^2\}+(\gamma^2-\epsilon)B^2-2\zeta(H,G_i) B+2z(H,G_i) A.\label{CK2}
\end{eqnarray} 
The Casimir operator $K$ can also be realized as a polynomial of the central elements  $H$ and $G_i$,
\begin{eqnarray}
K=\sum _k C_{ij} G_i^j H^k.\label{CK1}
\end{eqnarray} 
The algebraically independent integrals of motion of model (\ref{k1}) in parabolic coordinates are given by \cite{FH4}
\begin{eqnarray}
A=L^2+\frac{c_3 \xi}{4\eta}+\frac{c_2\eta}{4\xi},\quad
B=M_3+\frac{\xi-\eta}{\xi+\eta}\left\{\frac{c_3\xi^2-c_2\eta^2}{2\xi^2\eta-2\xi\eta^2}+\frac{c_0}{4}\right\},\label{k3}
\end{eqnarray}
where $\textbf{L}=\textbf{r}\times \textbf{p}-\frac{\textbf{r}}{r}Q$, $M_3=\frac{1}{2}\{(p_1L_2-p_2L_1)-(L_1p_2-L_2p_1)\}-\frac{z}{r}(a_1 H-\frac{c_1}{2}Q^2)$,
and for the model (\ref{KP1}) they are \cite{FH5}
\begin{eqnarray}
A= \sum_{j<k}{L}^2_{jk}+ \frac{2 r c_{1}}{(r+x_{0})} + \frac{2 r c_{2}}{(r-x_{0})},\quad
B=M_k+ \frac{c_{1}(r-x_{0})}{\hbar^{2} r(r+x_{0})}  + \frac{c_{2}(r+x_{0})}{\hbar^{2} r(r-x_{0})},\label{IntegralsAB}
\end{eqnarray}
where $L_{jk}=(x_j\pi_k-x_k\pi_j)-r^2\hbar F^a_{jk}T_a$, $j, k =0, 1, 2, 3, 4$, and $ M_{k}=\frac{1}{2}(\pi_jL_{jk}+L_{jk}\pi_j)+c_0\frac{x_k}{r}$ and $F^a_{jk}=\partial_j A^a_k-\partial_k A^a_j+\epsilon_{abc}A^b_j A^c_k$ is the Yang-Mills field tensor. $F^a_{jk}$ and $\pi_j=-i\hbar \frac{\partial}{\partial x_{j}} -\hbar A_{j}^{a} T_{a}$ obey the commutation relations
\begin{eqnarray}
[\pi_j, x_k]=-i\hbar\delta_{jk}, \quad [\pi_j,\pi_k]=i\hbar^2 F^a_{jk} T_a.
\end{eqnarray}

By direct computations and using various commutation identities, it was shown \cite{FH4, FH5} that the integrals of motion $\{A,B,C\}$ close to form the quadratic algebra $Q(3)$ (\ref{QA1}). For system (\ref{k1}), $Q$ and $L_3=\sqrt{\xi\eta}\cos\phi p_2-\sqrt{\xi\eta}\sin\phi p_1-\frac{\xi+\eta}{\xi-\eta}Q$ are central elements and they form the two dimensional Abelian algebra, 
 \begin{eqnarray}
[L_3,L_3]=0=[Q,Q].\label{LQ1}
\end{eqnarray} 
On the other hand, model (\ref{KP1}) has the first order integrals of motion $L_{jk}$, $j, k = 1, 2, 3, 4$ which generate the $so(4)$ Lie algebra.
Let
$\hat{L}^2=\sum_{j<k}{L}^2_{jk}$, $j, k = 1, 2, 3, 4$. Then
$\hat{L}^2$ is the Casimir operator of $so(4)$ and is also a central element of $Q(3)$. 
In the following table, we list the values of the structure constants in $Q(3)$ and $K$ for the models $H_K$ and $H_{YC}$, respectively,
\begin{table}[h] 
\begin{center}
    \begin{tabular}{ | l | p{6cm} | p{6cm} |}
    \hline
    &${\bf H_{K}}$ & ${\bf H_{YC}}$ 
    \\ \hline
   $\gamma$ & $2 $ & $2$ 
    \\ \hline
     $\epsilon$ &$c_2+c_3$& $8$
         \\ \hline
              $d$ & $8b_1H-4d_1Q^2-4c_4$ & $8H$
         \\ \hline
     $\zeta$ & $-4a_1HQL_3+2c_1Q^3L_3+c_0QL_3+a_1(c_2-c_3)H-\frac{1}{2}c_1(c_2-c_3)Q^2-\frac{1}{4}c_0(c_2-c_3)$ &  $-2c_0(c_1-c_2)-4c_0\hat{T}^2$
         \\ \hline
     $z$ & $2a^2H^2-2(ac_1+2b)HQ^2-4bHL_3^2+\frac{1}{2}(c_1^2+4d)Q^4+2dQ^2L_3^2+(4b-ac_0)H+\frac{1}{2}(c_0 c_1+4c_4-4d)Q^2+2c_4L_3^2+\frac{1}{8}(c_0^2-16c_4) $ &  $-4\hat{L}^2 H+(16-4c_1-4c_2)H+2c_o^2$
    \\
    \hline
    \end{tabular}
    \caption{\label{tabone} Coefficients in $Q(3)$ and $K$.}
\end{center}
\end{table}
\\
Thus both models $H_{K}$ and $H_{YC}$ have the quadratic algebra $Q(3)\oplus \mathcal{L}$ as their symmetry algebra, where $\mathcal{L}$ is $so(4)$ for $H_{YC}$ and $\mathcal{L}=0$ for $H_{K}$. Moreover, it was shown that the quadratic algebra $Q(3)$ for $H_{K}$ has structure constants involving $Q$, $L_3$, while $Q(3)\equiv Q(3; L^{so(4)}, T^{su(2)})$ for $H_{YC}$ with structure constants involving Casimir operators $\hat{L}^2$ of $so(4)$ and $\hat{T}^2$ of $su(2)$.

We now turn our attention to model (\ref{kf1}). The integrals of motion for this system were constructed via the recurrence approach \cite{FH6}. This approach was previously restricted to systems with scalar potentials. We first extended the application of this recurrence method to superintegrable monopole systems in \cite{FH6}. We here highlight the main results of \cite{FH6}.

The integrals of motion for the Schr\"{o}dinger St\"{a}ckel equivalent $H'_{MIC}\Psi=E'\Psi$ of $H_{MIC}\Psi=E\Psi$, where $E'=c_4+c_1\nu_2^2-2bE$ and $\Psi(r,\theta,\phi,\psi)=\chi(r,\theta)e^{i(\nu_1\phi+\nu_2\psi)}$, were constructed from the suitable combinations of the ladder and shift operators \cite{FH6},
\begin{eqnarray}
D_1=K^{+}_{l-\nu_2+\frac{1}{2},n}J^{+}_{l-\nu_1+1}J^{+}_{l-\nu_1} B, \quad D_2=BJ^{-}_{l-\nu_1}J^{-}_{l-\nu_1-1}K^{-}_{l-\nu_2+\frac{1}{2},n},
\end{eqnarray}
where $K^{\pm}_{l-\nu_2+\frac{1}{2},n}$ are the ladder operators, $J^{\pm}_{l-\nu_1}$ are the shift operators and $B=\sqrt{\textbf{L}^2+\frac{1}{4}}$ is a well-defined operator \cite{Lab1}.

The action of the combinations of operators $D_1D_2$ and $D_2D_1$ can be obtained on the wave functions. $D_1$ and $D_2$ form an algebraically independent set of differential operators and they close to form the higher-rank polynomial algebra $P(3)\oplus \mathcal{L}$ (see details in Ref. \cite{FH6}),
\begin{eqnarray}
&&[D_1, H'_{MIC}]=0=[D_2, H'_{MIC}],\quad [B, D_1]= 2D_1, \quad [B, D_2]=-2D_2,\nonumber
\\&& D_1D_2=f_1(H'_{MIC},B,L_3,Q),\quad D_2D_1=f_2(H'_{MIC},B,L_3,Q), \label{ff2}
\end{eqnarray}
where $f_1$ and $f_2$ are polynomials in the generators $H'_{MIC}$, $B$, $L_3$ and $Q$.

\section{Unirreps and spectra}
In this section, we focus on the finite dimensional unitary representations of the quadratic algebras and apply the results to derive the energy spectra for systems (\ref{k1}), (\ref{kf1}) and (\ref{KP1}).

The quadratic algebra $Q(3)$ can be realized in terms of the deformed oscillator algebra
\begin{eqnarray}
[\aleph,b^{\dagger}]=b^{\dagger},\quad [\aleph,b]=-b,\quad bb^{\dagger}=\Phi (\aleph+1),\quad b^{\dagger} b=\Phi(\aleph),\label{kpfh}
\end{eqnarray} 
with the structure function $\Phi(x)$ given by \cite{FH4, FH5}
\begin{eqnarray}
&\Phi(x; u,H)&=-3072\gamma^6 K\{2(x+u)-1\}^2+48d\gamma^8\{2(x+u)-3\}^2\nonumber\\&&\{2(x+u)-1\}^4\{2(x+u)+1\}^2+12288\gamma^4\zeta^2\nonumber\\&& +128\gamma^6(2\gamma z-d\epsilon)\{2(x+u)-1\}^2\{12(x+u)^2-12(x+u)-1\}\nonumber\\&&-256\gamma^4(2d\epsilon\gamma^2+12\epsilon\gamma z-4\gamma^3 z-3d\epsilon^2)\{2(x+u)-1\}^2.\label{ST1}
\end{eqnarray}
The corresponding values of the structure constants are given in Table 1 for models (\ref{k1}) and (\ref{KP1}), respectively.

On the other hand, for model (\ref{kf1}), the symmetry algebra (\ref{ff2}) was already in the form of the deformed oscillator algebra (\ref{kpfh}) by letting $\aleph=\frac{B}{2}$,  $b^{\dagger}=D_1$ and $b=D_2$. The structure function is given by \cite{FH6}
\begin{eqnarray}
&\Phi(x;u,H'_{MIC})&=\frac{(2x+u-2)}{256}[2(2x+u)-3]^2[2(2x+u)-2L_3-1]\nonumber\\&&[2(2x+u)-1] [2(2x+u)-2L_3-3][2(2x+u)-5]\nonumber\\&&[2(2x+u)-2L_3-4Q-3][H'_{MIC}-2\varepsilon\{(2x+u)-Q-1\}]\nonumber\\&&[2(2x+u)-4Q-1][H'_{MIC}+2\varepsilon\{(2x+u) -Q-1\}]\nonumber\\&&[2(2x+u)-4Q-3][2(2x+u)-2L_3-4Q-1].\label{ST2}
\end{eqnarray}

Using appropriate Fock space we can evaluate the structure functions (\ref{ST1}) and (\ref{ST2}) and obtain, for model (\ref{k1}), (\ref{KP1}) and the St\"{a}ckel equivalent of model (\ref{kf1}), respectively,
\begin{eqnarray}
\Phi(x;u, E_K)&=&-3145728 (c_4 - 2 bE_K + d q_2^2)[x+u-\tau_{-+}][x+u-\tau_{+-}]\nonumber\\&&[x+u-\tau_{--}] [x+u-\tau_{++}]\left[x+u-\left(\frac{1}{2}-\beta(E_K)\right)\right]\nonumber\\&&\left[x+u-\left(\frac{1}{2}+\beta(E_k)\right)\right],\label{pp1}
\end{eqnarray}
where $\tau_{\pm\pm}=1\pm m_1\pm m_2$, $m_1^2=c_2+(q_1-q_2)^2$, $m_2^2=c_3+(q_1+q_2)^2$, $\beta(E_K)=\frac{4aE_K-2c_1 q_2^2-c_0}{4\sqrt{c_4-2bE_K+dq_2^2}}$;
\begin{eqnarray}
&\Phi(x;u,E_{YC})&=[x+u-\mu_{++}][x+u-\mu_{+-}][x+u-\mu_{-+}][x+u-\mu_{--}]\nonumber\\&& \quad[ x+u -\eta(E_{YC})][ x+u-\eta(E_{YC})]  ,\label{pp2}
\end{eqnarray}
where $\mu_{\pm\pm}=\frac{1}{2}(1\pm n_1\pm n_2)$, $n_{1}^{2}=1+2c_{1} +\hbar^2 l_4(l_4+2) + 2 \hbar^2 T(T+1) $, $ n_{2}^{2}=1+2c_{2} +\hbar^2 l_4(l_4+2) - 2 \hbar^2 T(T+1)$, $\eta(E_{YC})=\frac{1}{2}+\frac{c_{0}}{\sqrt{-2E_{YC}}}$;
\begin{eqnarray}
&\Phi(x;u,E'_{MIC})&=\frac{(2x+u-2)}{256}[2(2x+u)-3]^2[2(2x+u)-2q_2-1]\nonumber\\&&[2(2x+u)-1] [2(2x+u)-2q_2-3][2(2x+u)-5]\nonumber\\&&[2(2x+u)-2q_2-4q_1-3][E'_{MIC}-2\varepsilon\{(2x+u)-q_1-1\}]\nonumber\\&&[2(2x+u)-4q_1-1][E'_{MIC}+2\varepsilon\{(2x+u)-q_1-1\}] \nonumber\\&&[2(2x+u)-4q_1-3][2(2x+u)-2q_2-4q_1-1].\label{pp3}
\end{eqnarray}
For finite dimensional unitary representations, we have the following constraints on the structure functions (\ref{pp1}), (\ref{pp2}) and (\ref{pp3}), 
\begin{equation}
\Phi(p+1; u,E)=0,\quad \Phi(0;u,E)=0,\quad \Phi(x)>0,\quad \forall x>0,\label{pro2}
\end{equation}
where $p$ is a positive integer. This gives $(p+1)$-dimensional unitary representations and their solution gives the corresponding energy $E_{K}$ \cite{FH4}, $E_{YC}$ \cite{FH5}, $E_{MIC}$ \cite{FH6},
for $\epsilon_1=\pm 1$, $\epsilon_2=\pm 1$,
\begin{eqnarray}
 &&2\beta(E_K)= 2 + 2 p  + \epsilon_1 m_1 + \epsilon_2 m_2;   
\\
&&E_{YC}=-\frac{c_{0}^{2}}{2( p+1+\frac{n_{1}+n_{2}}{2})^{2}};
\\&&
 E'_{MIC}=-\varepsilon(4p-2q_1+2\epsilon_2q_2+3).\label{en1}
\end{eqnarray}
By the coupling constant metamorphosis $E'\leftrightarrow c_4+c_1\nu_2^2-2bE$ and $\varepsilon^2\leftrightarrow \frac{c_0}{2}-2aE+d\nu_2^2$, we obtain the energy  of the original Hamiltonian (\ref{kf1}) from (\ref{en1})
\begin{eqnarray}
\frac{2bE_{MIC}-c_1q_2^2-c_4 }{\sqrt{\frac{c_0}{2}-2a E_{MIC}+dq_2^2}}=4p-2q_1+2\varepsilon_2 q_2+3.\label{en2}
\end{eqnarray}
The structure functions are positive for $\epsilon_1=1$, $\epsilon_2=1$ and $m_1, m_2>0$. The energies have degeneracy of $p+1$ only when these other quantum numbers would be fixed. The total number of degeneracies may be calculated by taking into account the further constraints on these quantum numbers.

{\bf Acknowledgements} The research of FH was supported by International Postgraduate Research Scholarship and Australian Postgraduate Award. IM was supported by the Australian Research Council, Discovery Project DP 160101376, YZZ was partially supported by the Australian Research Council, Discovery Project DP 140101492.

\section{Conclusion}
In this paper, we have reviewed  families of superintegrable Hamiltonian systems with Yang-Coulomb monopole interactions in the flat and curved Taub-NUT spaces \cite{FH4, FH5, FH6} and highlighted their unified algebra structures of the form $P(3)\oplus \mathcal{L}$. We have applied the direct and recurrence approaches to the superintegrable Hamiltonian systems interacting with monopoles and have given the algebraic derivations of their energy spectra. Our work represents the first application of the recurrence approach to models beyond scalar potentials such as the superintegrable monopole systems \cite{FH6}.
The direct and recurrence approaches could be extended to other $N$-dimensional MICZ-Kepler monopole type problems (e.g. \cite{Men1, Kri1}).

\end{document}